\documentclass[iop]{emulateapj}

\begin{document}

\shorttitle{Super-Eddington outburst of SMC X-3}

\title{{\it Swift} observations of SMC X-3 during its 2016-2017 super-Eddington outburst}

\author{Shan-Shan Weng$^{1}$, Ming-Yu Ge$^{2}$, Hai-Hui Zhao$^{1}$, Wei Wang$^{3,4}$, Shuang-Nan Zhang$^{2,4,5}$, Wei-Hao Bian$^{1}$, Qi-Rong Yuan$^{1}$}

\affil{$^{1}$ Department of Physics and Institute of Theoretical Physics,
Nanjing Normal University, Nanjing 210023, China}

\affil{$^{2}$ Key Laboratory of Particle Astrophysics, Institute of High Energy
Physics, Chinese Academy of Sciences, Beijing 100049, China}

\affil{$^{3}$ School of Physics and Technology, Wuhan University, 430072 Wuhan,
Hubei, China}

\affil{$^{4}$ National Astronomical Observatories, Chinese Academy of Sciences,
Beijing 100012, China}

\affil{$^{5}$ University of the Chinese Academy of Sciences, Beijing, China}

\email{wengss@njnu.edu.cn, gemy@ihep.ac.cn}
\begin{abstract}

The Be X-ray pulsar, SMC X-3 underwent a giant outburst from 2016 August to
2017 March, which was monitored with the {\it Swift} satellite. During the
outburst, its broadband flux increased dramatically, and the
unabsorbed X-ray luminosity reached an extreme value of $\sim 10^{39}$ erg/s
around August 24. Using the {\it Swift/XRT} data, we measure the observed pulse
frequency of the neutron star to compute the orbital parameters of the binary
system. After applying the orbital corrections to {\it Swift} observations, we
find that the spin frequency increases steadily from 128.02 mHz on August 10
and approach to spin equilibrium $\sim 128.74$ mHz in 2017 January
with an unabsorbed luminosity of $L_{\rm X} \sim 2\times10^{37}$ erg/s,
indicating a strong dipolar magnetic field $B \sim
6.8\times10^{12}$ G  at the neutron star surface. The spin-up rate is tightly correlated with its X-ray
luminosity during the super-Eddington outburst. The pulse profile in the {\it
Swift}/XRT data is variable, showing double peaks at the early stage of
outburst and then merging into a single peak at low luminosity. Additionally,
we report that a low temperature ($kT \sim 0.2$ keV) thermal component emerges
in the phase-averaged spectra as the flux decays, and it may be produced from
the outer truncated disk or the boundary layer between the exterior flow and
the magnetosphere.
\end{abstract}


\section{Introduction}

High-mass X-ray binaries contribute a large fraction of X-ray emission in
normal galaxies, and they are believed to reflect the recent star-formation
activities in their host galaxies \citep[e.g. ][]{grimm02,mineo12}. According to the
states of their optical companions, high-mass X-ray binaries can be subdivided
into supergiant X-ray binaries and Be/X-ray binaries (BeXBs). A BeXB consists of
a Be star and a compact object. Virtually, all confirmed compact objects in
BeXBs are neutron stars (NSs), and all these systems show X-ray pulsations (see
Reig 2011 for reviews). As young systems, NSs in BeXBs have high magnetic field
($B > 10^{10}$ G); therefore, BeXBs provide unique natural laboratories for
studying  physics in extremely strong gravity and magnetic fields.

The direct measurement of a NS magnetic field strength can be achieved from the
detection of a cyclotron scattering resonance feature \citep[e.g. ][ and
references therein]{coburn02, yan12, furst14, walter15}. Additionally,
investigating the interaction between the magnetosphere and the accretion
matter, we can acquire the information of NS magnetic field indirectly
\citep[e.g. ][]{weng11,shi15,christodoulou16}. That is, the effect of the
magnetic field strength manifest itself by the size of its magnetosphere
co-rotating with the central NS. The boundary of the magnetosphere is
determined where the ram pressure of in-falling flow is balanced by magnetic
pressure; thus, it expands with field strength and decreases with mass
accretion rate \citep{lamb73, ghosh79}. As the accretion rate decreases
below the critical value, the magnetospheric radius ($R_{\rm m}$) grows beyond
the corotation radius ($R_{\rm co} = \sqrt[3]{\frac{GMP^{2}}{4\pi^{2}}}$), at
which the Keplerian angular frequency is equal to the NS spin frequency, and
the centrifugal barrier spins away accretion matter. If most of material is
prevented from accreting onto NS, X-ray flux and pulsation decay sharply in a
few days, i.e. the ``propeller" effect \citep[e.g. ][]{cui97,campana14}.
Alternatively, the magnetosphere is penetrated into the corotation radius at
high luminosity, leading to the spin-up of a NS. If NSs are close to spin
equilibrium, their magnetic fields can be estimated from long-term averaged
spin parameters and X-ray luminosity \citep[e.g. ][]{klus14,shi15}. Meanwhile,
the torque reversals between steady spin-up and spin-down are commonly shown in
BeXBs \citep[e.g. ][]{bildsten97}.

Besides the long-term average spin evolution, the instantaneous torque
measurements during episodic outbursts are essential to test accretion torque
theories. Transient BeXBs experience periodic and less energetic ($L_{\rm X} <
10^{37}$ erg/s) outbursts or rare giant outbursts, which are referred to as
type I and type II outbursts, respectively \citep{reig11}. The tight
relationships between spin-up rate and (pulsed) flux detected in the luminous
outbursts of BeXBs (e.g. A0535+262 and 2S 1417-624) are interpreted as the sign
of transient accretion disks around NSs, which are supported by the detection
of simultaneous quasi-periodic oscillations (QPOs) \citep{finger96,sartore15}.
It is worth to note that a small number of sources (e.g. SMC X-1, LMC X-4, 4U
0115+63, V0332+53) can reach a peak X-ray luminosity in excess of $10^{38}$
erg/s \citep[e.g. ][ and references therein]{li11,mushtukov15b}. Intriguingly,
a growing number of ultraluminous X-ray sources (ULXs) in nearby galaxies
have been found to exhibit coherent pulsations
\citep{bachetti14,furst16,israel16,israel17}, indicating a connection to X-ray
pulsars \citep{shao15, mushtukov15b, kawashima16, king16, mushtukov17}.
Nowadays, super-Eddington accretion in magnetized NSs draw more attention
\citep[e.g. ][]{eksi15,pan16,tsygankov16,chen17}. However, a detailed study on
such dramatic phenomena is hampered by the lack of observations.

The Small Magellanic Cloud (SMC) is the second nearest galaxy \citep[$d = 62.1$
kpc; ][]{hilditch05,graczyk14, scowcroft16} after the Large Magellanic Cloud,
and it has high-mass X-ray binaries in abundance due to the recent star-forming
activities \citep{zaritsky02,sturm13,yang17}. SMC X-3 (also known as SXP 7.78)
was discovered with {\it SAS 3} X-ray observatory in 1978 \citep{clark78}, and
was identified as an accreting pulsar with a detected pulsation of 7.78 second
\citep{edge04}. The spectral type of the optical counterpart is identified as
B1--B1.5 with $V = 14.91$ \citep{mcbride08}. The orbital period $\sim 44.9$
days was detected in both X-ray and optical bands \citep{corbet03, cowley04,
galache08,bird12}. However, its eccentricity and other orbital parameters
are still unknown. Recently, SMC X-3 underwent a giant type II outburst in 2016
with a peak X-ray luminosity of $\sim 10^{39}$ erg/s, and it was monitored in
Target of Opportunity mode by {\it Swift} since 2016 August 10. On 2016
November 8, we reported our preliminary results on analyzing the {\it Swift}
data \citep{weng16}, which are the basis of this work. During preparation of
this manuscript, \citet[][ submitted]{townsend17} investigated the optical and
X-ray data (including the {\it Swift} data) of SMC X-3, and obtained
similar orbital parameters of the binary as given in our paper. In this paper,
we focus on the {\it Swift} data and carry out a comprehensive analysis on
these data to investigate the physics of super-Eddington accretion around a
magnetized NS. The data reduction is described in the next section. In Section
3, we perform the timing analysis and calculate the orbital elements of the
binary. In Section 4, we discuss the physical implications of these results and
present our main conclusions.

\section{Data Reduction}

The {\it MAXI}/GSC was triggered by brightening of SMC X-3 on 2016 August 8
\citep{negoro16}, which was confirmed by {\it Swift} during its survey of the
SMC \citep{kennea16}. In this paper, we analyze all {\it Swift} pointing
observations taken between 2016 August 10 and 2017 January 1. The {\it Swift}
Gamma Ray Burst Explorer carries three scientific instruments covering a broad
energy range of $\sim 0.002-150$ keV: the Burst Alter Telescope (BAT), the
X-ray Telescope (XRT), and the UV/Optical Telescope (UVOT) \citep{gehrels04}.
The BAT daily light curve is adopted from \cite{krimm13}
\footnote{\protect\url{http://swift.gsfc.nasa.gov/results/transients/}}.
Meanwhile, both the XRT and the UVOT data are processed with the packages and
tools available in \textsc{heasoft} 6.19.

\begin{figure}
\begin{center}
\includegraphics[scale=0.45]{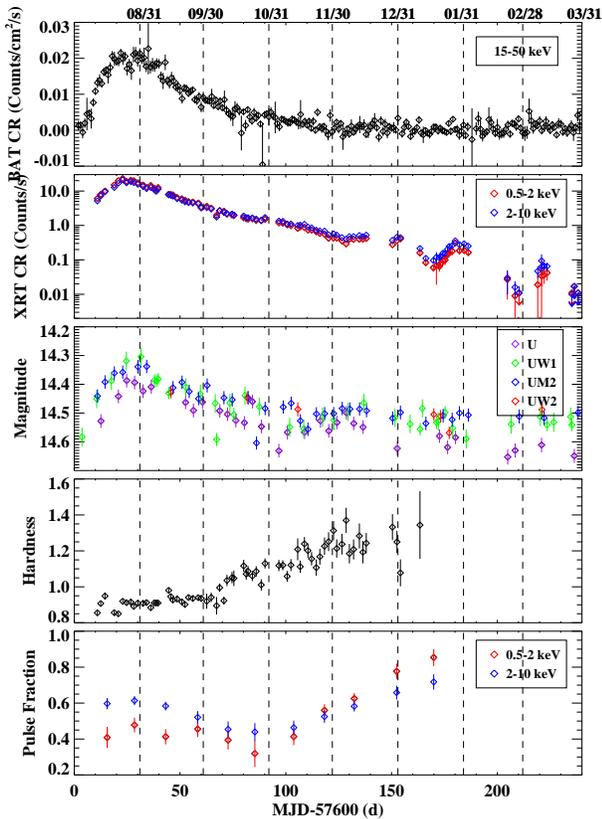}
\caption{Panels from top to bottom show the light curves from the BAT, XRT,
UVOT data, the evolution of the 2--10 keV to 0.5--2 keV hardness
ratios, and the pulse fractions, respectively. The hardness ratios after MJD
57765 are not shown in the bottom panel due to large uncertainties. \label{lc}}
\end{center}
\end{figure}

When the source is bright, the observations are carried out in the
windowed-timing (WT) mode. While after 2017 January 16, the count rate in
0.5--10 keV is less than 0.5 cts/s, and the observations are performed in the
photon-counting (PC) mode without significant pile-up effects (Table
\ref{log}). For the XRT data, the initial event cleaning is executed with the
task \texttt{xrtpipeline} using standard quality cuts. The source and
background events are extracted from a circle and an annulus region centered at
the source position, respectively. The light curves are corrected for the
telescope vignetting and point-spread-function losses with the task
\texttt{xrtlccorr}, and then are subtracted by the scaled background count rate
to yield the net light curves (Figure \ref{lc}). The hardness for each
observation is calculated as the ratio of average count rates (CR) in
(2.0--10.0 keV)/(0.5--2.0 keV) bands. The source becomes undetected in
last three observations, the upper limit of count rates are estimated by using the
X-ray image package XIMAGE.

The ancillary response files are created using the task \texttt{xrtmkarf} and
the latest response files are taken from the CALDB database for the spectral
analyses. The spectral fitting is restricted to the 0.6--10 keV energy range
due to calibration residuals below 0.6 keV for the WT mode data\footnote{see
http://swift.gsfc.nasa.gov/docs/heasarc/caldb/swift/docs/xrt/
SWIFTXRT-CALDB-09.pdf}. For the PC mode data, we employ the C-statistic
\citep{cash79} instead of common $\chi^{2}$ for spectral fitting in 0.3--10 keV
because of low count rates and short exposure times. Unfortunately, spectra are
unavailable for some observations which have very limited photons (Table
\ref{log}). The absorption column density along the line of sight to SMC X-3 is
difficult to constrain and could yield an extremely small value ($nH < 10^{14}$
cm$^{-2}$) for some of the observations; therefore, we fix it to the Galactic
absorption towards the direction of the source \citep[$6.57\times10^{20}$
cm$^{-2}$, ][]{dickey90}. At the early stage of the outburst, the
phase-averaged spectra can be well fitted by an absorbed power-law (PL) model
with a photon index of $\sim 0.7-1.2$. These results are consistent with those
seen in {\it NuSTAR} data, which can be fitted by a cutoff PL model with a
photon index $\sim 0.5-0.7$ and a folding energy of $\sim 11-15$ keV
\citep{pottschmidt16,tsygankov17}. During 2016 September 25 and December 12,
the soft excess emerges below 1 keV, which can be described by a black-body
(BB) emission with $kT \sim 0.1-0.2$ keV and $R_{\rm BB} \sim 10^2-10^3$ km.
The soft thermal component is variable and contributes a small fraction ($\sim
1-3$\%) of total flux in 0.6--10 keV; however, because of the low S/N ratio of
band-limited data, we cannot put a tight constraint on this component.

\begin{figure}
\begin{center}
\includegraphics[scale=0.55]{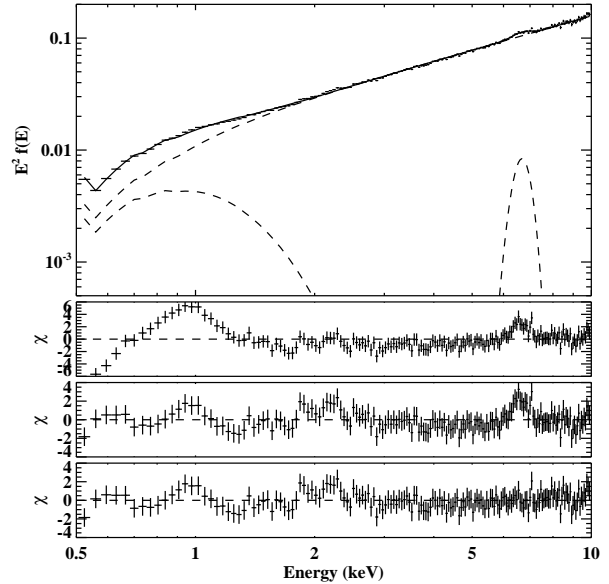}
\caption{Fitting to the XMM-Newton EPIC-pn spectrum observed on 2016 October
14-15 (ObsID = 0793182901). Panels from top to bottom show the unfolded spectra
using the (BB+PL+Gau) model, the fitting residuals for the a single PL model,
the (BB+PL) model, and with an additional Gaussian line. \label{spec}}
\end{center}
\end{figure}

In order to rule out the instrumental influence, we also analyze the
XMM-Newton EPIC-pn spectrum observed on 2016 October 14-15 and confirm the
existence of the thermal component. The EPIC-MOS data were taken in imaging
mode and suffered from the pile-up effect; therefore, we only use the EPIC-pn
data which was in timing mode. The data in the first 4.5 kilo seconds are
excluded because of background flares, and the spectrum is extracted from the
rest data with an exposure time of 28 kilo seconds. The EPIC-pn spectrum cannot
be well fitted by a single PL model with a reduced $\chi^2$ larger than 2.9. When a
cool BB component is added, the reduced $\chi^2$ decreases to 1.07, and it
further reduces to 0.75 with including an additional Gaussian line
(phabs*(bbodyrad+powerlaw+gauss) in XSPEC, Figure \ref{spec}). The best fitted
parameters are: $nH = 1.4_{-0.2}^{+0.2}\times10^{21}$ cm$^{-2}$, $kT =
0.19_{-0.01}^{+0.01}$ keV, $Norm_{\rm BB} = 1136_{-392}^{+560}$, $\Gamma =
0.99_{-0.01}^{+0.01}$, $Norm_{\rm PL} = 1.54_{-0.03}^{+0.03}\times10^{-2}$,
$E_{\rm l} = 6.65 _{-0.10}^{+0.10}$ keV, $\sigma = 0.35_{-0.10}^{+0.14}$ keV,
$Norm_{\rm Gau} = 1.64_{-0.45}^{+0.53}\times10^{-5}$, and $\chi^2/dof =
123.8/165$. More detailed analysis on the XMM-Newton data will be presented else where.

\begin{deluxetable*}{ccccccc}
\tabletypesize{\tiny} \tablewidth{0pt} \tablecaption{Log of {\it Swift}/XRT
data} \tablehead{\colhead{ObsID} & \colhead{Date} & \colhead{Exposure}
&\colhead{Mode} & \colhead{$R_{\rm Source}$} &
\colhead{$R_{\rm Background}$} & \colhead{$L_{\rm 0.6-10 keV}$} \\
\colhead{} & \colhead{} & \colhead{(second)} & \colhead{} & \colhead{(pixel)} &
\colhead{(pixel/pixel)} & \colhead{(10$^{37}$ erg/s)}} \startdata \hline
00034673001&  2016-08-10&     4671&  WT  & 25 & 25/50  & $35.3_{-0.4}^{+0.4}$ \\
00034673002&  2016-08-12&     1978&  WT  & 25 & 25/50  & $46.9_{-0.7}^{+0.7}$ \\
00034673003&  2016-08-14&      935&  WT  & 25 & 25/50  & $59.0_{-1.1}^{+1.1}$ \\
00034673004&  2016-08-18&     1997&  WT  & 25 & 25/50  & $68.3_{-0.8}^{+0.8}$ \\
00034673005&  2016-08-20&     2050&  WT  & 25 & 25/50  & $84.0_{-1.0}^{+1.0}$ \\
00034673006&  2016-08-22&     1958&  WT  & 25 & 25/50  & $90.1_{-0.9}^{+0.9}$ \\
00034673007&  2016-08-24&     1291&  WT  & 25 & 25/50  & $102.7_{-1.3}^{+1.4}$ \\
00034673008&  2016-08-26&      504&  WT  & 25 & 25/50  & $98.2_{-1.9}^{+1.9}$ \\
00034673009&  2016-08-28&     1583&  WT  & 25 & 25/50  & $98.2_{-1.1}^{+1.1}$ \\
00034673010&  2016-08-30&      953&  WT  & 25 & 25/50  & $88.1_{-1.3}^{+1.3}$ \\
00034673011&  2016-09-01&      793&  WT  & 25 & 25/50  & $92.9_{-1.6}^{+1.6}$ \\
00034673012&  2016-09-03&     1932&  WT  & 25 & 25/50  & $78.9_{-1.0}^{+1.0}$ \\
00034673013&  2016-09-05&     1345&  WT  & 25 & 25/50  & $56.6_{-1.0}^{+1.0}$ \\
00034673014&  2016-09-06&     2985&  WT  & 25 & 25/50  & $61.1_{-0.6}^{+0.6}$ \\
00034673015&  2016-09-07&     2977&  WT  & 25 & 25/50  & $61.3_{-0.7}^{+0.7}$ \\
00034673016&  2016-09-08&     2764&  WT  & 25 & 25/50  & $57.2_{-0.7}^{+0.7}$ \\
00034673017&  2016-09-13&     2973&  WT  & 25 & 25/50  & $45.1_{-0.5}^{+0.5}$ \\
00034673018&  2016-09-14&     2979&  WT  & 25 & 25/50  & $44.2_{-0.5}^{+0.5}$ \\
00034673019&  2016-09-15&     2981&  WT  & 25 & 25/50  & $42.8_{-0.5}^{+0.5}$ \\
00034673020&  2016-09-17&     3004&  WT  & 25 & 25/50  & $33.5_{-0.4}^{+0.4}$ \\
00034673021&  2016-09-19&     2918&  WT  & 25 & 25/50  & $30.8_{-0.4}^{+0.4}$ \\
00034673022&  2016-09-21&     2591&  WT  & 25 & 25/50  & $27.4_{-0.4}^{+0.4}$ \\
00034673023&  2016-09-23&     2992&  WT  & 25 & 25/50  & $26.0_{-0.4}^{+0.4}$ \\
00034673024&  2016-09-25&     2610&  WT  & 25 & 25/50  & $24.9_{-0.4}^{+0.4}$ \\
00034673025&  2016-09-27&     3289&  WT  & 25 & 25/50  & $23.6_{-0.4}^{+0.4}$ \\
00034673026&  2016-09-29&     2969&  WT  & 25 & 25/50  & $19.3_{-0.4}^{+0.4}$ \\
00034673027&  2016-10-01&      553&  WT  & 25 & 25/50  & $17.6_{-0.7}^{+0.8}$ \\
00034673028&  2016-10-03&     1718&  WT  & 25 & 25/50  & $16.2_{-0.4}^{+0.4}$ \\
00034673029&  2016-10-06&      726&  WT  & 25 & 25/50  & $17.5_{-0.9}^{+1.0}$ \\
00034673030&  2016-10-07&     2789&  WT  & 25 & 25/50  & $14.9_{-0.3}^{+0.3}$ \\
00034673031&  2016-10-09&     2839&  WT  & 25 & 25/50  & $13.2_{-0.3}^{+0.3}$ \\
00034673032&  2016-10-11&     1982&  WT  & 25 & 25/50  & $12.6_{-0.3}^{+0.4}$ \\
00034673033&  2016-10-13&     2028&  WT  & 25 & 25/50  & $12.2_{-0.3}^{+0.3}$ \\
00034673034&  2016-10-14&     1199&  WT  & 25 & 25/50  & $11.6_{-0.4}^{+0.5}$ \\
00034673035&  2016-10-19&     2493&  WT  & 25 & 25/50  & $10.2_{-0.3}^{+0.3}$ \\
00034673036&  2016-10-20&     2852&  WT  & 25 & 25/50  & $9.62_{-0.26}^{+0.26}$ \\
00034673037&  2016-10-21&     3435&  WT  & 25 & 25/50  & $9.31_{-0.22}^{+0.23}$ \\
00034673038&  2016-10-23&     3685&  WT  & 25 & 25/50  & $9.37_{-0.22}^{+0.22}$ \\
00034673039&  2016-10-25&     3626&  WT  & 25 & 25/50  & $8.51_{-0.21}^{+0.22}$ \\
00034673040&  2016-10-27&     2959&  WT  & 25 & 25/50  & $7.67_{-0.25}^{+0.25}$ \\
00034673041&  2016-10-29&     3984&  WT  & 25 & 25/50  & $9.04_{-0.23}^{+0.23}$ \\
00034673042&  2016-11-06&     5256&  WT  & 25 & 25/50  & $7.16_{-0.17}^{+0.18}$ \\
00034673043&  2016-11-04&     3973&  WT  & 25 & 25/50  & $7.59_{-0.22}^{+0.22}$ \\
00034673044&  2016-11-08&     4646&  WT  & 25 & 25/50  & $6.57_{-0.18}^{+0.18}$ \\
00034673045&  2016-11-10&     4453&  WT  & 25 & 25/50  & $6.30_{-0.18}^{+0.18}$ \\
00088012001&  2016-11-13&     1803&  WT  & 25 & 25/50  & $5.56_{-0.26}^{+0.26}$ \\
00034673046&  2016-11-14&     4452&  WT  & 25 & 25/50  & $5.78_{-0.16}^{+0.16}$ \\
00034673047&  2016-11-16&     4328&  WT  & 25 & 25/50  & $5.91_{-0.19}^{+0.19}$ \\
00034673048&  2016-11-18&     4628&  WT  & 25 & 25/50  & $5.04_{-0.15}^{+0.16}$ \\
00034673049&  2016-11-20&     4881&  WT  & 25 & 25/50  & $5.07_{-0.16}^{+0.16}$ \\
00034673050&  2016-11-22&     3287&  WT  & 25 & 25/50  & $4.21_{-0.17}^{+0.17}$ \\
00034673051&  2016-11-24&     4484&  WT  & 25 & 25/50  & $3.89_{-0.14}^{+0.14}$ \\
00034673053&  2016-11-26&     4430&  WT  & 25 & 25/50  & $3.82_{-0.15}^{+0.15}$ \\
00034673054&  2016-11-28&     4589&  WT  & 25 & 25/50  & $3.37_{-0.14}^{+0.15}$ \\
00034673055&  2016-11-30&     5026&  WT  & 25 & 25/50  & $3.35_{-0.12}^{+0.13}$ \\
00034673056&  2016-12-02&     4563&  WT  & 25 & 25/50  & $3.01_{-0.13}^{+0.13}$

\enddata
\tablecomments{$R_{\rm Source}$: Radius of source region; $R_{\rm Background}$:
Inner and outer radius of background region; $L_{\rm 0.6-10 keV}$: Unabsorbed
luminosities are calculated with the distance of $d =$ 62.1 kpc. \label{log}}
\end{deluxetable*}

\setcounter{table}{0}

\begin{deluxetable*}{ccccccc}
\tabletypesize{\tiny} \tablewidth{0pt} \tablecaption{Log of {\it Swift}/XRT
data} \tablehead{\colhead{ObsID} & \colhead{Date} & \colhead{Exposure}
&\colhead{Mode} & \colhead{$R_{\rm Source}$} &
\colhead{$R_{\rm Background}$} & \colhead{$L_{\rm 0.6-10 keV}$} \\
\colhead{} & \colhead{} & \colhead{(second)} & \colhead{} & \colhead{(pixel)} &
\colhead{(pixel/pixel)} & \colhead{(10$^{37}$ erg/s)}} \startdata \hline
00034673057&  2016-12-04&     4915&  WT  & 25 & 25/50  & $2.57_{-0.12}^{+0.12}$ \\
00034673058&  2016-12-06&     4917&  WT  & 25 & 25/50  & $2.24_{-0.11}^{+0.11}$ \\
00034673059&  2016-12-08&     4526&  WT  & 25 & 25/50  & $2.68_{-0.12}^{+0.13}$ \\
00034673060&  2016-12-10&     4711&  WT  & 25 & 25/50  & $2.70_{-0.11}^{+0.11}$ \\
00034673061&  2016-12-12&     3073&  WT  & 25 & 25/50  & $2.92_{-0.16}^{+0.16}$ \\
00034673062&  2016-12-14&     3538&  WT  & 25 & 25/50  & $2.83_{-0.13}^{+0.13}$ \\
00034673063&  2016-12-16&     4380&  WT  & 25 & 25/50  & $3.08_{-0.12}^{+0.13}$ \\
00034673064&  2016-12-28&     4189&  WT  & 25 & 25/50  & $2.25_{-0.11}^{+0.11}$ \\
00034673065&  2016-12-30&     4288&  WT  & 25 & 25/50  & $2.36_{-0.12}^{+0.12}$ \\
00034673066&  2017-01-01&     2026&  WT  & 25 & 25/50  & $2.30_{-0.18}^{+0.19}$ \\
00034673067&  2017-01-10&     1179&  WT  & 25 & 25/50  & $1.37_{-0.23}^{+0.26}$ \\
00034673069&  2017-01-13&     1666&  WT  & 25 & 25/50  & $0.75_{-0.11}^{+0.12}$ \\
00034673071&  2017-01-16&      768&  PC  & 15 & 15/30  & $0.68_{-0.08}^{+0.09}$ \\
00034673073&  2017-01-18&       82&  PC  & 15 & 15/30  & \nodata$^{\S}$ \\
00034673074&  2017-01-19&      445&  PC  & 15 & 15/30  & $0.77_{-0.11}^{+0.12}$ \\
00034673075&  2017-01-20&      382&  PC  & 15 & 15/30  & $0.83_{-0.11}^{+0.12}$ \\
00034673076&  2017-01-21&      347&  PC  & 15 & 15/30  & $1.05_{-0.14}^{+0.16}$ \\
00034673077&  2017-01-22&      329&  PC  & 15 & 15/30  & $1.06_{-0.15}^{+0.17}$ \\
00034673078&  2017-01-23&      355&  PC  & 15 & 15/30  & $1.38_{-0.15}^{+0.17}$ \\
00034673079&  2017-01-24&      336&  PC  & 15 & 15/30  & $1.67_{-0.17}^{+0.19}$ \\
00034673080&  2017-01-25&      235&  PC  & 15 & 15/30  & $1.65_{-0.19}^{+0.22}$ \\
00034673081&  2017-01-27&     2295&  WT  & 15 & 15/30  & $1.98_{-0.16}^{+0.16}$ \\
00034673082&  2017-01-29&      237&  PC  & 15 & 15/30  & $1.75_{-0.20}^{+0.23}$ \\
00034673083&  2017-01-31&      394&  PC  & 15 & 15/30  & $1.86_{-0.16}^{+0.17}$ \\
00034673084&  2017-02-02&      552&  PC  & 15 & 15/30  & $1.76_{-0.14}^{+0.15}$ \\
00034673087&  2017-02-20&      138&  PC  & 15 & 15/30  & \nodata$^{\S}$ \\
00034673088&  2017-02-24&      385&  PC  & 15 & 15/30  & \nodata$^{\S}$ \\
00034673089&  2017-02-26&      411&  PC  & 15 & 15/30  & \nodata$^{\S}$ \\
00034673091&  2017-03-07&      198&  PC  & 15 & 15/30  & \nodata$^{\S}$ \\
00034673092&  2017-03-08&       57&  PC  & 15 & 15/30  & \nodata$^{\S}$ \\
00034673093&  2017-03-09&      345&  PC  & 15 & 15/30  & $0.46_{-0.10}^{+0.13}$ \\
00034673094&  2017-03-10&      384&  PC  & 15 & 15/30  & $0.46_{-0.10}^{+0.12}$ \\
00034673095&  2017-03-11&      350&  PC  & 15 & 15/30  & $0.42_{-0.08}^{+0.09}$ \\
00034673096&  2017-03-23&      229&  PC  & 15 & 15/30  & undetected \\
00034673097&  2017-03-24&      104&  PC  & 15 & 15/30  & undetected \\
00034673098&  2017-03-26&      192&  PC  & 15 & 15/30  & undetected
\tablecomments{$\S$: Spectra are unavailable due to short exposure times and
low flux.}
\end{deluxetable*}

In addition to the pointing data, we include the 27 survey observations (on the
SMC) with an average exposure of $\sim 60$ s for the following photometry. For
each observation, we first sum the sky images with \texttt{uvotimsum}, and then
perform aperture photometry with the summed images by using
\texttt{uvotsource}. A source aperture of radius 5 arcsec and a larger
neighbouring source-free region for background are used. The AB magnitudes for
all filters are shown in Figure \ref{lc}.

\section{Timing Analysis \& Results}

\begin{figure}
\begin{center}
\includegraphics[scale=0.45]{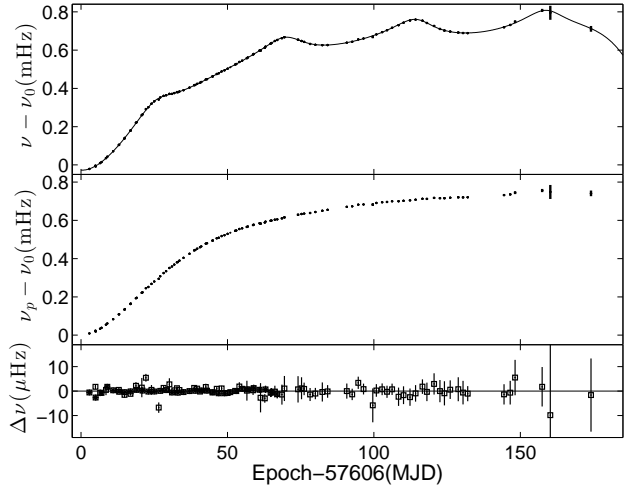}\caption{The spin frequencies without the
orbital correction are plotted versus time (begin time: MJD 57606) in the upper
panel($\nu_{0}=0.128$Hz), while the frequencies shown in the middle panel have been
orbitally corrected. The bottom panel shows fitting residuals of the spin and
orbital parameters. The filled square data are adopted from the {\it Fermi}/GBM
results. \label{freq}}
\end{center}
\end{figure}

The spin evolution during giant outbursts is crucial for investigating the
accretion processes in BeXBs and for determining orbital parameters by using
the orbital {\it Doppler} effect \citep[e.g. ][]{li11}. Only WT mode
data, which have a high time resolution and relatively long exposure time for
individual observations (Table \ref{log}), are used for the following timing
analysis. Before computing the spin frequency, we apply the barycentric
correction with the ftool \texttt{barycorr} to the source event files (in the
range of 0.5--10 keV). The source position is adopted from the 2MASS
all sky Catalog of point sources \citep{cutri03}. The spin frequency is
obtained by folding the observed counts to reach the maximum Pearson $\chi^2$.
The derived values of the observed spin frequency are the same as that
presented by the {\it Fermi}/GBM Pulsar
Project\footnote{\protect\url{http://gammaray.nsstc.nasa.gov/gbm/science/pulsars.html}};
however, the source becomes faint and below the detection limit of GBM after
MJD 57671 (Figure \ref{freq}).

\begin{figure}
\begin{center}
\includegraphics[scale=0.45]{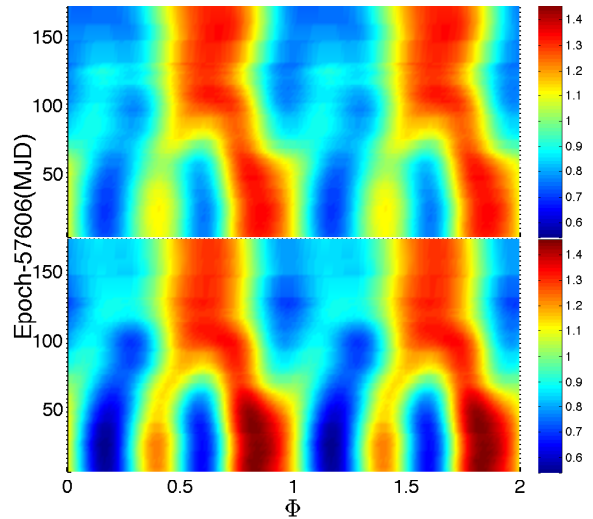}\caption{Evolution of the background-substracted pulse
profiles in 0.5-2 keV (upper panel) and 2-10 keV bands (bottom panel). The
color bar marks the normalized CR. \label{phase}}
\end{center}
\end{figure}

The background-subtracted light curves with a time bin size of 0.05 s
are extracted from the barycentric corrected event files, and are folded over
the best frequency. For the purpose of comparing the profiles between
observations, the pulse profiles are aligned by the cross correlation function
(Figure \ref{phase}). The energy-dependent pulse profiles make
clear that the spectra harden during the peak. The pulse fraction is defined as
$PF=(M-N)/(M+N)$, where $M$ and $N$ are the maximum and minimum flux of the profile,
respectively. To reduce the statistical error, we calculate the mean
pulse fraction in every 15 days, which are shown in the bottom panel of Figure
\ref{lc}.

\subsection{Orbital Elements}

To study the frequency evolution, the {\it Doppler} effects of the binary
should be considered here. However, the frequency evolution is very fast, so we
utilize seven frequency derivatives to describe spin evolution. The detailed
processes are described as follows: (1)The spin frequencies are searched
without considering the spin evolution; (2) Fit the spin parameters and orbital
elements; (3) Search spin frequencies considering the spin evolution and
orbital modulation; (4) Repeat steps (2) and (3) for several times to get the
best spin parameters and orbital elements. The fitting in step (2) is based on
the Levenberg-Marquardt algorithm for nonlinear least square method, weighted
by the error of each point, which is similar to the fitting process described
in Li et al. (2011, and references therein). The spin frequency errors of GBM
are multiplied by a factor of 5 to balance the weights between XRT and GBM in
the fitting process. The best fitted parameters listed in Table
\ref{table_timing_para} are quite consistent with those reported in
\citet{townsend17} and the uncertainties of parameters is 1\,$\sigma$.

\begin{figure}
\begin{center}
\includegraphics[scale=0.5]{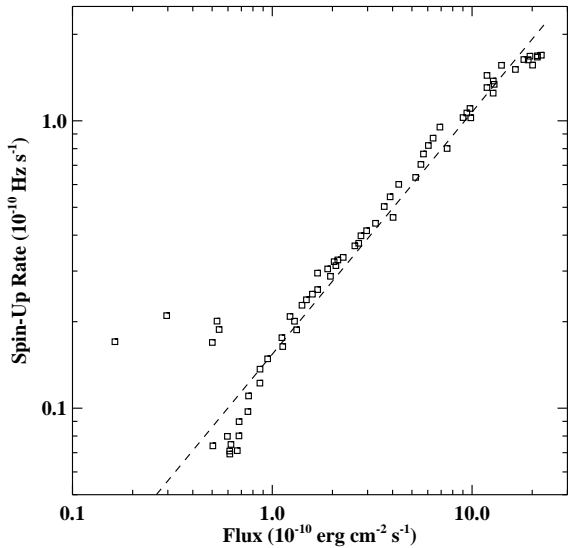}\caption{Relationship between the 0.6-10
keV unabsorbed flux and the spin-up rate during the 2016 super-Eddington
outburst. The dashed line is the best-fit power-law with an index of
0.84$\pm$0.02. Note that, the spin-up rates for the five data points well above
the dashed line have large uncertainties and are consistent with zero within
2$\sigma$. \label{flux_spin}}
\end{center}
\end{figure}

It is difficult to obtain the derivative of the spin frequency with each individual
observation alone. Alternatively, we calculate the spin-up rates from the fitting
results and the X-ray fluxes for each two successive observations, and plot
them in Figure \ref{flux_spin}.

\begin{table}
\caption{Timing results of SMC X-3 from the X-ray Observations}
\scriptsize{} \label{table_timing_para}
\begin{center}
\begin{tabular}{l l l }
\hline\hline
    &  Parameters              & Value                                                        \\
\hline
    & R.A.                     & $00^{h}52^{m}05^{s}.64$                                      \\
    & Decl.                    & -72$^{\textordmasculine}$26$^{\prime}$04$^{\prime\prime}$.2  \\
\hline
    & Epoch(MJD)                         & 57606                                              \\
    & $\nu$(mHz)                         & 128.005(2)                                       \\
    & $\dot{\nu}$(10$^{-5}$Hz d$^{-1}$)  & -0.10(5)                                          \\
    & $\ddot{\nu}$(10$^{-6}$Hz d$^{-2})$ & 2.08(7)                                           \\
    & $\nu_{3}$(10$^{-7}$Hz d$^{-3})$    &-1.88(7)                                           \\
    & $\nu_{4}$(10$^{-8}$Hz d$^{-4})$    & 1.1(4)                                           \\
    & $\nu_{5}$(10$^{-10}$Hz d$^{-5})$   & -3.9(2)                                          \\
    & $\nu_{6}$(10$^{-12}$Hz d$^{-6})$   &  8.9(5)                                          \\
    & $\nu_{7}$(10$^{-13}$Hz d$^{-7})$   & -0.96(7)                                          \\
    & Porb(d)                            & 44.52(9)                                            \\
    & asini(light-second)                & 194(1)                                             \\
    & e                                  & 0.259(3)                                           \\
    & $\omega{\textordmasculine}$        & 202(2)                                            \\
    & $T_{\omega}$(MJD)                  & 57632.1(3)                                         \\
    & $\chi^{2}_{re}$(MJD)               & 1.1                                         \\
    & d.o.f.                             & 89                                         \\
\hline
\end{tabular}
\end{center}
\end{table}

\section{Discussion \& Conclusions}

During the 2016-2017 giant outburst, the X-ray luminosity of SMC X-3 reached a
peak value $\sim 10^{39}$ erg/s (in 0.6--10 keV) around August 24, that is a
few times of NS Eddington luminosity and close to the low-luminosity tail of
ULXs. Since 2017 January, the source flux dropped sharply with two humps
separated by about one orbital period ($\sim 42$ d), and becomes lower than the
detection limit of {\it Swift}/XRT (Figure \ref{lc}). Its X-ray spectra become
harder as flux decays. The BAT daily light curve shows an indication of flat
plateau during MJD $\sim 57618-57640$. The magnitudes in the near-ultraviolet
bands increased by $\sim 0.2^{m}-0.3^{m}$ during outburst, and returned to a
constant level after two months. Alternatively, the U band flux changes with
the XRT count rates with the Spearman's rank correlation coefficients of
$\rho/P = 0.86/3.6\times10^{-10}$.

According to the {\it Doppler} motion of the binary, we fit the orbital
modulation of the observed pulse period, obtain an orbital period of $P =
44.52\pm0.09$ days that is consistent with the previous works to within
3$\sigma$ uncertainty \citep{galache08, bird12}, and determine the projected
semi-major axis $asini=194\pm1$ light seconds and an eccentricity of \textbf{$e
= 0.259\pm0.003$.} After correcting for the orbital motion, we find that the
spin frequency increases steadily, indicating the formation of a transient
accretion disk surrounding the central NS during the outburst. According to the
accretion torque theory, we would expect the spin-up rate increases with the
accretion rate ($\dot{\nu} \propto \dot{M}^{6/7}$) \citep{ghosh79,wang81}. The
power-law relationships between $\dot{\nu}$ and X-ray luminosity have been
confirmed in some BeXBs during their type II giant outbursts and consistent
with the results of QPOs evolution \citep[e.g.
][]{finger96,icdeml11,sartore15}. However, the fitted power-law indices are
generally greater than the prediction of 6/7. Alternatively, fitting the
spin-up rate and the 0.6--10 keV flux of SMC X-3 with a power-law, we obtain an
index of ${0.84\pm0.02}$ that is in agreement with 6/7, but the relation deviates
from the power-law at the peak and the low luminosity (Figure \ref{flux_spin}).

\cite{klus14} investigated all {\it RXTE}/PCA data to determine the long-term
average spin period \textbf{($P = 7.7836\pm0.0001$ s)}, spin-down rate
($\dot{P} = 0.00262\pm0.00003$ s yr$^{-1}$), and the average X-ray luminosity
($L_{\rm X} \sim 3.7\times10^{36}$ erg/s) for SMC X-3. These results point to a
torque switch from spin-down to spin-up at the beginning of the giant outburst
($L_{\rm X} < 10^{38}$ erg/s). By assuming spin equilibrium, they derived a
magnetic field of $\sim 2.9\times10^{12}$ G. Here, we report the instantaneous
spin period measurements during the 2016 giant outburst. The spin frequency
with the orbital correction is evidently near the torque equilibrium with the
spin frequency derivative close to zero at an unabsorbed luminosity of $\sim
L_{\rm X} \sim 2\times10^{37}$ erg/s assuming the distance of 62.1 kpc (Figres
\ref{freq} and \ref{flux_spin}). Thus, we can estimate a magnetic field of SMC
X-3 with the assumption of $R_{\rm co} = R_{\rm m}$, or $B = [4.8\times10^{10}
P^{7/6}(\frac{flux}{10^{-9}\ {\rm erg\ cm^{-2}\
s^{-1}}})^{1/2}\times(\frac{d}{1\ {\rm kpc}})\times(\frac{M}{1.4\
M_{\odot}})^{1/3}\times (\frac{R}{10^{6}\ {\rm cm}})^{-5/2}]$ G for the
simplified model \citep{cui97}, yielding $B \sim 6.8\times10^{12}$ G with the
canonical value of NS mass and radius, i.e. 1.4 $M_{\odot}$ and 10 km. That is,
the obtained value is significantly higher than that reported by \cite{klus14}.

Intriguingly, if taking the color correction factor into account, the emission
size of the thermal component ($R_{\rm BB} \sim 1000$ km) detected during the
source returning to the quiescence state is significantly larger than the NS
radius. Alternatively, the spin periodic modulation indicates that the BB
emission is not homogenous, but comes from the place even farther away from the
central NS. The location where the thermal component is produced could be close
to the corotation radius of SMC X-3 ($R_{\rm co} \sim 6000$ km), supporting the
torque equilibrium hypothesis. The emission line detected in XMM-Newton data is
consistent with that from the highly ionized Fe XXV, which could originate from
the illumination of cold material (i.e. the cool BB component) by central hard
X-rays. The relatively large line widths ($\sigma =
0.35_{-0.10}^{+0.14}$ keV) can be interpreted as the results of Keplerian
rotation at a radius of $\sim 1000$ km, which agrees with the size of the BB
component discussed above. We, however, caution that the broad emission line
with the central energy of 6.65 keV could be due to the blending of lines from
different ionisation states of Fe, i.e. Fe K$\alpha$ and Fe K$\beta$, which
have been detected in other accreting pulsars \citep[e.g. ][]{reynolds10}.

After 2017 January, the flux drops abruptly, which gives the hint of
``propeller" effect as the result of the further decrease in accretion rate and that
the centrifugal force prevents material from entering the magnetosphere. It is
worth to note that there are two X-ray flux jumps exhibited around MJD 57780
and 57822 (corresponding to the orbital phase of $\sim 0.3$) in Figure
\ref{lc}, and they can be interpreted as the increase in mass accretion rate
when the NS travels through its periastron, i.e. Type I outbursts. Since the
typical X-ray luminosity of its Type I outburst is above the {\it Swift}/XRT
detection limit, we expect that the periodic outbursts can be observed by the
future {\it Swift} monitoring data. However, these outbursts are beyond the
scope of this paper, and we suggest that the source has ended its 2016-2017
giant outburst in the end of 2017 March.

At low luminosity, the single peak pulse profiles were observed in the {\it
RXTE} \citep{galache08}, {\it Chandra} and {\it XMM-Newton} archival data
\citep{haberl08} more than ten years ago. As shown explicitly in Figure
\ref{phase}, the pulse profiles of SMC X-3 during the 2016 outburst are
variable and have double peaks at the beginning of the outburst, and then
converges to a single peak as the flux decays below $L_{\rm X} \sim
4\times10^{37}$ erg/s. The similar evolution pattern of pulse profiles has been
reported in other outbursts but with lower transient luminosity, e.g. the 1994
giant outburst of A0535+262 \citep{bildsten97}. These results can be
interpreted as different geometries of the accretion column at beyond and below
the critical luminosity \citep[$\sim 10^{37}$ erg/s; e.g. ][]{basko76,
becker12, mushtukov15a, sartore15}, respectively. At lower luminosity, the
in-falling gas may be decelerated via Coulomb interactions and a pencil-beam of
emission is formed. At the supercritical state, the accretion flow is
decelerated via the radiative shock and the photons can only escape from
accretion column walls resulting in double peaks structure \citep[i.e. fan beam
mode; e.g. ][]{becker12}. As can be seen in Figure \ref{lc}, the pulse fraction
generally increases with the hardness during the outburst, which is consistent
with those recently found in ULXs. Investigating a sample of ULX broadband
X-ray spectra, \cite{pintore17} suggested that the ULX pulsars have harder
spectra than that of the majority of other ULXs and the spectra became softer
when no pulsations are detected. It is worth to note that the luminosity of all
three accreting NSs can be two orders of magnitude higher than that of SMC X-3
but all have single peak pulse profiles \citep{bachetti14, furst16, israel16,
israel17}. More extreme surface magnetic fields are required to account for the
observed characteristics of ULX pulsars.

\acknowledgments{We thank the anonymous referee for his/her helpful comments.
S.S.W. thanks Dr. Kim Page from {\it Swift} team for discussions on data
analysis. We thank Jian Li, Xiao-Chuan Jiang and Fang-Jun Lu for many valuable
suggestions. This work is supported by the National Natural Science Foundation
of China under grants 11303022, 11133002, 11233001, 11573023, 11173016,
11673013, 11373024, 11233003, 11503027, 11622326 and 11433005, National Program
on Key Research and Development Project (Grant No. 2016YFA0400802 and 2016YFA0400803), and by the
Special Research Fund for the Doctoral Program of Higher Education (grant No.
20133207110006).}

\end{document}